\documentclass[reqno]{amsart}
\usepackage{amssymb}
\usepackage{hyperref}
\usepackage{graphicx}
\usepackage{booktabs}
\usepackage{multirow}

\newtheorem{theorem}{Theorem}[section]
\newtheorem{proposition}[theorem]{Proposition}
\newtheorem{lemma}[theorem]{Lemma}
\newtheorem{corollary}[theorem]{Corollary}

\theoremstyle{definition}

\theoremstyle{remark}
\newtheorem{remark}{Remark}[section]
\newtheorem*{note}{Note}

\numberwithin{equation}{section}

\newcommand{\tbin}[2]{\genfrac{[}{]}{0pt}{}{#1}{#2}}

\begin{document}

\title[Gaudin's determinantal norm formula]
{Gaudin's determinantal norm formula and Hall--Littlewood polynomials}

\author{Jan Felipe  van Diejen}

\address{
Instituto de Matem\'aticas, Universidad de Talca,
Casilla 747, Talca, Chile}

\email{diejen@utalca.cl}


\thanks{This work was supported in part by the Agencia Nacional de Investigaci\'on y Desarrollo (ANID), through
FONDECYT Grant \# 1250427.}


\begin{abstract}
The Hall--Littlewood polynomials are orthogonal with respect to a discrete measure stemming from a discretization of the Lieb--Liniger model that describes $q$-bosons on the circle. This note establishes a previously conjectured Gaudin-type determinantal formula for their quadratic norms.
\end{abstract}

\maketitle

\section{Introduction}\label{sec1}

The Lieb--Liniger model consists of $N$ spinless bosons on a circle that interact pairwise by means of a repulsive `delta potential' \cite{mat:many-body}; its spectral problem was
solved in the seminal work \cite{lie:exact} through a coordinate Bethe Ansatz, which computed the
energy spectrum in terms of the roots of a corresponding algebraic system of Bethe-Ansatz equations. A fundamental observation due to Yang and Yang revealed that any such Bethe root can be interpreted as the absolute minimum of an associated strictly convex Morse function
\cite{yan-yan:thermodynamics}.
Finally, the orthogonality and the completeness of the Lieb--Liniger Bethe-Ansatz eigenfunctions were established by Dorlas in \cite{dor:orthogonality}.

In \cite{gau:modeles}---a precursor of his influential 1983 monograph \cite{gau:bethe}---Gaudin proposed
a beautiful closed formula for the quadratic norm of the Bethe-Ansatz eigenfunction in terms of the determinant of the Hessian of the
Yang--Yang Morse function evaluated at the pertinent Bethe root (cf.~\cite[\S 4.3.3]{gau:bethe} for an English account).
A major breakthrough was achieved by Korepin
\cite{kor:calculations}, who proved Gaudin's conjectural norm formula by means of powerful new tools tailored to the algebraic Bethe Ansatz.
Refinements and generalizations of Korepin's approach constitute the current state of the art in the computation of
scalar products between eigenstates of quantum integrable models that are exactly solvable within the framework of the algebraic Bethe Ansatz \cite{kor:quantum,sla:algebraic}.

The aim of the present note is to establish a Gaudin-type determinantal norm formula for the Hall--Littlewood polynomials,
with respect to a finite-dimensional discrete orthogonality measure stemming from a discretization of the Lieb--Liniger model introduced
in \cite{die:diagonalization}.  The pertinent norm formula was first presented as a conjecture in
 \cite{die:finite-dimensional} (and verified for $N\leq 4$ and for $t=0$),
 where it was shown to imply the Gaudin--Korepin norm formula via a straightforward continuum limit used in \cite{die:diagonalization}
 to give an elementary proof of the orthogonality of the Lieb--Liniger Bethe-Ansatz eigenbasis.
In \cite{kor:cylindric}, Korff observed that in the $N$-particle sector, the spectral problem
for the $q$-boson model in its Hall--Littlewood formulation \cite{tsi:quantum} is equivalent to that of the discrete Lieb--Liniger model from \cite{die:diagonalization}, with the deformation parameter of the $q$-bosons playing the role of the Hall--Littlewood parameter $t$ (we follow Macdonald's notation \cite{mac:symmetric}, so the letter $q$ appears here only in the names of the model and of standard basic hypergeometric identities). By conjugation of partitions (interchanging rows and columns), the discrete Lieb--Liniger/$q$-boson model can also be interpreted as a quantum relativistic Toda chain
on the periodic integer lattice \cite{die:spectrum,duv-pas:q-bosons}.
The upshot is that our Gaudin-type determinantal norm formula for the Hall--Littlewood polynomials establishes the quadratic norm formulae for the Bethe-Ansatz eigenfunctions of both the discrete Lieb--Liniger/$q$-boson model and the quantum relativistic Toda chain on the periodic integer lattice, and---via the continuum limit---also leads to a novel proof of the celebrated Gaudin--Korepin norm formula. 

The material is organized as follows. Section \ref{sec2} sets up the notation and states our finite-dimensional orthogonality relation for the Hall--Littlewood polynomials
as the main result in Theorem \ref{affine-HL-orthogonality:thm}. The orthogonality relation in question is supported on the Bethe roots of the discrete Lieb--Liniger model and constitutes an affine analog of the standard orthogonality relation for the Hall--Littlewood polynomials in terms of a torus integral against Macdonald's orthogonality weight function.
The corresponding Gaudin-type quadratic norm formula is subsequently deduced as the diagonal part of the dual orthogonality relation in Corollary \ref{dual-affine-HL-orthogonality:cor}.
The proofs are detailed in Section \ref{sec3}.
The approach differs substantially from Korepin's algebraic Bethe Ansatz method. In a nutshell, viewing the Bethe roots as preimages of lattice points under the gradient map of the Yang--Yang Morse function (the \emph{Bethe map}), we use Poisson summation to convert the sum over the Bethe roots into a sum of torus integrals of Hall--Littlewood polynomials against twisted versions of Macdonald's orthogonality weight. Only finitely many of these integrals are actually nonzero. These nonzero contributions are evaluated by the standard orthogonality relation for the Hall--Littlewood polynomials in fewer variables and subsequently collected into a product formula by means of the $q$-Chu--Vandermonde summation formula.

\begin{remark}
The discrete orthogonality relations for the Hall--Littlewood polynomials studied in \cite{die:diagonalization,die:finite-dimensional} fit within a much wider scheme of orthogonality
relations for multivariate polynomials stemming from the construction of Gaussian-type cubature rules \cite{xu:minimal}. From this perspective,
the Gaudin-type determinantal norm formula for the Hall--Littlewood polynomials proved in this note permits the computation
of the Christoffel weights of a cubature rule for the integration of homogeneous symmetric functions with respect to the density of the circular unitary ensemble, cf.~\cite[Section 2]{die:cubature}. At $t=0$ the Hall--Littlewood polynomials simplify to Schur polynomials. The cubature rule then reduces
to a previously known one on the fundamental alcove of the affine Weyl group of $\mathfrak{sl}(N,\mathbb{C})$ \cite{li-xu:discrete,moo:cubature} (cf.\ Remark \ref{t=0:rem} below).
\end{remark}

\begin{remark}
The idea underlying our approach towards the Gaudin-type norm formula is rooted in the Berry--Tabor trace formula for
integrable quantum systems \cite{ber-tab:closed}, in which the sum over the quantum numbers
labeling the spectrum is Poisson-resummed into integrals over the action variables: the term
with vanishing dual index gives the smooth part of the level density, while the remaining terms
produce the oscillatory contributions of the closed classical orbits. This program was implemented
recently, at a formal level, for the Lieb--Liniger model \cite{urb-kel-ric:periodic}, where the quantization conditions
are the Bethe equations and the relevant change of variables is the Bethe map
from rapidities to Bethe quantum numbers, whose Jacobian is the Gaudin matrix governing the
norms of the Bethe states. The same change of variables had been employed earlier by
Pozsgay \cite{poz:contributions} and Woynarovich \cite{woy:normalization} in their analysis of
partition functions of Bethe Ansatz systems in the thermodynamic limit, where only the term with
vanishing dual index of the Poisson summation is retained (cf.\ \cite[App.~E]{woy:normalization}). In the present
finite-dimensional setting the Bethe map is the gradient of the Yang--Yang Morse function, and
the Poisson summation becomes an exact Fourier expansion on the torus in which only finitely
many dual terms survive: the term with vanishing dual index reproduces the standard
orthogonality relation of the Hall--Littlewood polynomials with respect to Macdonald's weight,
while the remaining terms---the counterparts of the oscillatory periodic-orbit corrections in
the trace formula---account for the affine modification of the dual orthogonality weights.
In view of the equivalence pointed out in \cite{kor:cylindric}, our results establish rigorously for the $q$-boson model---albeit by a route that
bypasses the algebraic Bethe Ansatz---the Gaudin-type determinantal formula for the squared norms
of the Bethe states that was deduced in \cite[Section 2.2]{oku-yos:gg} by formally applying the derivation of
on-shell inner products due to Slavnov (based on the commutation relations of the Yang--Baxter
algebra) \cite{sla:algebraic}.
\end{remark}

\begin{note}
Following standard conventions, $\mathbb{N}$ and $ \mathbb{N}_0$ denote the \emph{positive} integers and the \emph{nonnegative} integers, respectively.
Moreover, throughout we will use the convention that empty products are equal to $1$ and empty sums are equal to $0$.
\end{note}

\section{Main result}\label{sec2}

\subsection{Hall--Littlewood polynomials}
For a partition $\lambda=(\lambda_1,\lambda_2,\dots)$, i.e., a weakly decreasing sequence of nonnegative integers with finitely many positive terms (the parts),
we denote its \emph{weight} by $|\lambda|:=\sum_{j\in\mathbb{N}} \lambda_j$ and its \emph{length} by $\ell(\lambda):=|\{j\in\mathbb{N} \mid\lambda_j>0\}|$ (where $| \cdot |$ counts the cardinality of the set in question). Let us write $\Lambda$ for the set of all partitions.

For $c,k\in\mathbb{N}_0$, we write $\Lambda^{(k)}:=\{\lambda\in\Lambda\mid\ell(\lambda)\le k\}$ for the partitions of length at most $k$, and $c^k:=(\underbrace{c,\dots,c}_{k\ \text{parts}})$ for the \emph{rectangular partitions}.
Given a pair of positive integers $n$, $m$, let us set $\boxed{N:=n+1}$ and define \emph{bounded partitions}
\begin{equation}
\Lambda^{(n,m)}:=\{\lambda\in\Lambda^{(n)}\mid\lambda\subset m^n\} .
\end{equation}
Here $\lambda\subset\mu$ denotes the \emph{inclusion relation for partitions} ($\lambda_j\le\mu_j$ for all $j\in\mathbb{N}$), so the parts of $\lambda$ are bounded by those of the rectangular partition $m^n$.

For $\lambda\in\Lambda^{(N)}$, the \emph{Hall--Littlewood polynomial} is a symmetric polynomial in the variables $z:=(z_1,\ldots,z_N)$ of the form \cite[Chapter III]{mac:symmetric}:
\begin{subequations}
\begin{equation}\label{HL}
R_\lambda(z)=R_\lambda(z;t) :=\sum_{\sigma\in S_N}C(z_{\sigma(1)},\dots,z_{\sigma(N)};t)\,z_{\sigma(1)}^{\lambda_1}\cdots z_{\sigma(N)}^{\lambda_N},
\end{equation}
where the summation is over the group $S_N$ of all permutations $\sigma=\bigl(\begin{smallmatrix}1&\cdots&N\\ \sigma(1)&\cdots&\sigma(N)\end{smallmatrix}\bigr)$ of $\{1,\ldots ,N\}$, and
\begin{equation}\label{C}
C(z)=C(z_1,\ldots, z_N;t) := \prod_{1\le j<k\le N}\frac{1-t z_j^{-1}z_k}{1-z_j^{-1}z_k}\quad \text{with}\quad -1<t<1.
\end{equation}
\end{subequations}

The following orthogonality relation for the Hall--Littlewood polynomials is well known and appears in the literature in many equivalent guises. In its present formulation,
the orthogonality relation can be conveniently retrieved from, e.g., \cite[Theorem 2]{ven:vanishing}:
\begin{subequations}
\begin{equation}\label{HL-orthogonality}
\forall\lambda,\kappa\in\Lambda^{(N)}:\qquad \frac{1}{(2\pi)^N\, N!}\int_{[0,2\pi]^N}R_\lambda (e^{\mathrm{i}\xi} )\,
R_\kappa (e^{-\mathrm{i}\xi} )\,\Delta (e^{\mathrm{i}\xi} )\,\mathrm{d}\xi=\frac{\delta_{\lambda,\kappa}}{\hat{\Delta}_\lambda} ,
\end{equation}
where $e^{\mathrm{i}\xi} := \bigl( e^{\mathrm{i}\xi_1},\ldots ,e^{\mathrm{i}\xi_N}\bigr)$, $\mathrm{d}\xi:=\mathrm{d}\xi_1\cdots\mathrm{d}\xi_N$, $\delta_{\lambda,\kappa}$ refers to the Kronecker delta symbol, while
\begin{equation}\label{wfs}
 \Delta(z):=\frac{1}{C(z)C(z^{-1})}\quad\text{and}\quad \hat{\Delta}_\lambda :=  \frac{1}{ \prod_{i\in\mathbb{N}_0} [\mathrm{m}_i(\lambda)]!} .
\end{equation}
\end{subequations}
Notice that here our count of the multiplicities of the parts of partitions $\lambda\in\Lambda^{(N)}$ includes $\mathrm{m}_0(\lambda)=N-\ell(\lambda)$ $(\geq 0)$, i.e.,
\begin{equation}
\forall i\in\mathbb{N}_0:\quad \mathrm{m}_i (\lambda):=|\{1\le j\le N\mid\lambda_j=i\}| ,
\end{equation}
and we have employed $t$-deformed integers and factorials of the standard geometric-series form:
\begin{equation}
\forall k\in\mathbb{N}_0:\quad [k]:=\sum_{0\leq l <k} t^l=\frac{1-t^k}{1-t}\quad\text{and}\quad  [k]!:=\prod_{1\leq l \leq k} [l].
\end{equation}

\subsection{Morse function, Hessian, and critical points}

For any $\mu\in\Lambda^{(n,m)}$, we define the following smooth function $V_\mu:\mathbb{R}^N\to \mathbb{R}$ of the variable $\xi=(\xi_1,\ldots,\xi_N)$:
\begin{subequations}
\begin{equation}\label{Morse}
V_\mu (\xi):=\sum_{1\le j<k\le N}\int_0^{\xi_j-\xi_k}v (\theta)\,\mathrm{d}\theta+\sum_{1\le j\le N}\Bigl(\frac m2\xi_j^2-2\pi\bigl(\rho_j+\mu_j-\tfrac{|\mu|}{N}\bigr)\xi_j\Bigr) ,
\end{equation}
where
\begin{equation}\label{v}
v (\theta)=v (\theta;t):=\int_0^\theta \frac{1-t^2}{1+t^2-2t\cos\vartheta}\, \mathrm{d}\vartheta\quad\text{and}\quad \rho_j:=\frac{n}{2} +1-j .
\end{equation}
\end{subequations}
The equations for the critical points of $V_\mu (\xi)$ are given by the system
\begin{equation}\label{crit}
m\xi_j+\sum_{\substack{1\leq k\leq N\\ k\neq j}}v(\xi_j-\xi_k)=2\pi\bigl(\rho_j+\mu_j-\tfrac{|\mu |}N\bigr)\quad\text{for}\quad 1\leq j\leq N ,
\end{equation}
and the matrix elements $H_{jk}(\xi) = \frac{\partial^2V_\mu(\xi)}{\partial \xi_j\partial\xi_k}$ of the Hessian read explicitly:
\begin{subequations}
\begin{equation}\label{Hess}
H_{jk}(\xi)=\Bigl(m+\sum_{1\leq l\leq N}v'(\xi_j-\xi_l)\Bigr)\delta_{jk}- v'(\xi_j-\xi_k)\quad \text{for}\quad  1\leq j,k\leq N ,
\end{equation}
with
\begin{equation}\label{vprime}
  v'(\vartheta)=\frac{1-t^2}{1+t^2-2t\cos\vartheta}>0 .
\end{equation}
\end{subequations}

\begin{proposition}[{\cite[Section 4]{die:diagonalization}}]
The function $V_\mu (\xi)$ \eqref{Morse}, \eqref{v} is a smooth Morse function that is both \emph{(i)} strictly convex and \emph{(ii)} radially unbounded: $V_\mu(\xi)\to +\infty$ when
$\langle\xi,\xi\rangle\to\infty$ (where $\langle\cdot ,\cdot\rangle$ denotes the standard inner product on $\mathbb{R}^N$).
\end{proposition}

Consequently, the system \eqref{crit} for the critical points has a unique solution $\xi=\xi_\mu$ given by the global minimizer of the Morse function $V_\mu (\xi)$ \eqref{Morse}, \eqref{v}.

\begin{remark}\label{hyperplane:rem}
It is immediate from \eqref{crit} that the critical point $\xi_\mu$ lies in the hyperplane
$\mathbb{R}_0^N:=\{\xi\in\mathbb{R}^N\mid \xi_1+\cdots+\xi_N=0\} $, because the right-hand sides add up to zero and $\sum_{1\le j\neq k\le N} v(\xi_j-\xi_k)=0$ since $v(\theta)$ is an odd function; hence $m(\xi_1+\cdots+\xi_N)=0$.
\end{remark}

\subsection{Affine orthogonality relation for the Hall--Littlewood polynomials}
Let $\xi_\mu$, $\mu\in\Lambda^{(n,m)}$, denote the unique global minimum of the strictly convex Morse function $V_\mu (\xi)$ \eqref{Morse}, \eqref{v}, and let $H(\xi)$ denote the positive definite
Hessian matrix with elements given by \eqref{Hess}, \eqref{vprime}.

The following theorem provides a level-$m$ affine counterpart of the orthogonality relation \eqref{HL-orthogonality}, \eqref{wfs} for the Hall--Littlewood polynomials.
To formulate it, we count the multiplicities of the parts of $\lambda\in\Lambda^{(n,m)}$ modulo $m$:
\begin{equation}\label{mult-affine}
\forall i\in\mathbb{Z}/m\mathbb{Z}:\quad \mathrm{m}^{\mathrm{aff}}_i (\lambda):=|\{1\le j\le N\mid\lambda_j\equiv i \pmod{m}\}| ,
\end{equation}
so $\mathrm{m}^{\mathrm{aff}}_0(\lambda)=\mathrm{m}_0(\lambda)+\mathrm{m}_m(\lambda)$ and $\mathrm{m}^{\mathrm{aff}}_i(\lambda)=\mathrm{m}_i(\lambda)$ for $0<i<m$.

\begin{theorem}[Affine Orthogonality Relation]\label{affine-HL-orthogonality:thm}
The Hall--Littlewood polynomials $R_\lambda (z;t)$ \eqref{HL}, \eqref{C} satisfy the following orthogonality relation
\begin{subequations}
\begin{equation}\label{orthogonality-affine}
\forall \lambda,\kappa\in \Lambda^{(n,m)}:  \qquad \sum_{\nu\in\Lambda^{(n,m)}} R_\lambda( e^{\mathrm{i} \xi_\nu})   R_\kappa( e^{-\mathrm{i} \xi_\nu}) \Delta^{\mathrm{aff}} (\xi_\nu) =\frac{\delta_{\lambda,\kappa}}{\hat{\Delta}^{\mathrm{aff}}_\lambda}  ,
\end{equation}
with
\begin{equation}\label{wf-affine}
 \Delta^{\mathrm{aff}} (\xi )  :=\frac{m\,\Delta(e^{\mathrm{i}\xi})}{N\det H(\xi)}
\end{equation}
and
\begin{equation}\label{dual-wf-affine}
\hat{\Delta}^{\mathrm{aff}}_\lambda :=  \frac{1}{ \prod_{i\in\mathbb{Z}/m\mathbb{Z}} [\mathrm{m}^{\mathrm{aff}}_i(\lambda)]!}
=\frac{1}{ [\mathrm{m}_0(\lambda)+\mathrm{m}_m(\lambda)]! \prod_{0<i<m} [\mathrm{m}_i(\lambda)]!} .
\end{equation}
\end{subequations}
\end{theorem}

Notice that $\hat{\Delta}^{\mathrm{aff}}_\lambda=\hat{\Delta}_\lambda$ \eqref{wfs} for all $\lambda\in\Lambda^{(n,m)}$ with $\lambda_1<m$. Moreover, after rewriting the products directly
in terms of the parts of the partition as follows
\begin{equation}\label{dual-wf-affine-parts}
\hat{\Delta}^{\mathrm{aff}}_\lambda=\prod_{\substack{1\leq j<k\leq N\\ \lambda_j=\lambda_k}}\frac{1-t^{k-j}}{1-t^{1+k-j}}
\prod_{\substack{1\leq j<k\leq N\\ \lambda_j=\lambda_k+m}}\frac{1-t^{n+1-(k-j)}}{1-t^{n+2-(k-j)}} ,
\end{equation}
it becomes manifest that the weights \eqref{wf-affine}, \eqref{dual-wf-affine} are precisely those of \cite[Eqs.~(3.3c), (3.4a)]{die:finite-dimensional}. Indeed, the first product in \eqref{dual-wf-affine-parts} runs over the blocks of equal parts and yields $\prod_{i\in\mathbb{N}_0}1/[\mathrm{m}_i(\lambda)]!$, whereas the second product runs over the $\mathrm{m}_m(\lambda)\,\mathrm{m}_0(\lambda)$ pairs $(j,k)$ with $\lambda_j=m$ and $\lambda_k=0$, and gives rise to the factor $[\mathrm{m}_0(\lambda)]!\,[\mathrm{m}_m(\lambda)]!/[\mathrm{m}_0(\lambda)+\mathrm{m}_m(\lambda)]!$.

The corresponding dual orthogonality relation establishes the Gaudin-type norm formulae conjectured in \cite[Conjecture 2]{die:finite-dimensional}.

\begin{corollary}[Dual Affine Orthogonality Relation]\label{dual-affine-HL-orthogonality:cor}
The Hall--Littlewood polynomials $R_\lambda (z;t)$ \eqref{HL}, \eqref{C} satisfy the following dual orthogonality relation
\begin{equation}\label{dual-orthogonality-affine}
\forall \mu,\nu\in \Lambda^{(n,m)}:  \qquad\sum_{\lambda\in\Lambda^{(n,m)}} R_\lambda( e^{\mathrm{i} \xi_\mu})   R_\lambda( e^{-\mathrm{i} \xi_\nu}) \hat{\Delta}^{\mathrm{aff}}_\lambda
 =\frac{\delta_{\mu,\nu}}{\Delta^{\mathrm{aff}} (\xi_\mu)} .
\end{equation}
\end{corollary}

\begin{remark}\label{unitary:rem}
Since $t$ is real, $\overline{R_\lambda (e^{\mathrm{i}\xi} )}=R_\lambda (e^{-\mathrm{i}\xi} )$ and $\overline{C(e^{\mathrm{i} \xi})}=C(e^{-\mathrm{i} \xi})$, so the weights
$\Delta(e^{\mathrm{i}\xi})=|C(e^{\mathrm{i}\xi})|^{-2}$ and $\Delta^{\mathrm{aff}}(\xi_\nu)$ are real and both orthogonality relations are Hermitian. More precisely, $\Delta(e^{\mathrm{i}\xi})\geq 0$, with equality iff $\xi_j\equiv\xi_k \pmod{2\pi}$ for some $j\neq k$, and $\Delta^{\mathrm{aff}}(\xi_\nu)>0$ for all $\nu\in\Lambda^{(n,m)}$, because the coordinates of the nodes $\xi_\nu$ are pairwise distinct modulo $2\pi$ (cf.\ Lemma \ref{Phi:lem} (iii) and Eq.~\eqref{node} below, as $\delta+\nu$ is $L$-regular).
Thus \eqref{HL-orthogonality}, \eqref{wfs} states that the Hall--Littlewood polynomials $R_\lambda(e^{\mathrm{i}\xi})$, $\lambda\in\Lambda^{(N)}$, form an orthogonal system in $L^2\bigl([0,2\pi]^N,\Delta(e^{\mathrm{i}\xi})\mathrm{d}\xi\bigr)$, while Theorem \ref{affine-HL-orthogonality:thm} states that the $|\Lambda^{(n,m)}|$ functions $\nu\mapsto R_\lambda(e^{\mathrm{i}\xi_\nu})$, $\lambda\in\Lambda^{(n,m)}$, form an orthogonal basis of the $|\Lambda^{(n,m)}|$-dimensional Hilbert space $\ell^2\bigl(\Lambda^{(n,m)},\Delta^{\mathrm{aff}}\bigr)$. In other words, \eqref{orthogonality-affine} guarantees that the rows of the matrix
\begin{equation*}
\Bigl[\, R_\lambda(e^{\mathrm{i}\xi_\nu})\sqrt{\Delta^{\mathrm{aff}}(\xi_\nu)\,\hat{\Delta}^{\mathrm{aff}}_\lambda}\ \Bigr]_{\lambda,\nu\in\Lambda^{(n,m)}} 
\end{equation*}
are orthonormal, i.e., this matrix is unitary and its columns are therefore orthonormal too, which is precisely the dual orthogonality relation of Corollary \ref{dual-affine-HL-orthogonality:cor} (and implies,
in particular, that the nodes $\xi_\nu$, $\nu\in\Lambda^{(n,m)}$, are pairwise distinct). This confirms that Corollary \ref{dual-affine-HL-orthogonality:cor} is a direct consequence
of Theorem \ref{affine-HL-orthogonality:thm}.
\end{remark}

\begin{remark}
Corollary \ref{dual-affine-HL-orthogonality:cor} provides the quadratic norms of the Bethe-Ansatz eigenbases for the
discrete Lieb--Liniger model in \cite[Section 5]{die:diagonalization} and for the (equivalent) quantum relativistic Toda chain on the periodic integer lattice in \cite[Section 4]{die:spectrum}.
In the former reference the particles are labeled by $1,\ldots ,n$ rather than by $1,\ldots ,N$, so the $n$ there corresponds to our $N=n+1$ (whereas $m$ and $t$ keep their meaning). In the latter reference the parameter $t$ is denoted by $q$ and the roles of $m$ and $n$ are interchanged, while the wave functions live on the conjugate partitions $\mu\in\Lambda^{(m,n)}$ via $\psi_{\xi_\nu}(\mu)=R_{\mu'}(e^{\mathrm{i}\xi_\nu})$, and the weights of the inner product are normalized as multinomials $[N]!\,\hat{\Delta}^{\mathrm{aff}}_{\mu'}$ rather than $\hat{\Delta}^{\mathrm{aff}}_{\mu'}$, so the quadratic norms there are $[N]!/\Delta^{\mathrm{aff}}(\xi_\nu)$.
\end{remark}

\section{Proof of the affine orthogonality relation}\label{sec3}
Throughout this section $\langle\cdot ,\cdot\rangle$ denotes the standard inner product on $\mathbb{R}^N$, $|x|:=x_1+\cdots +x_N$ for $x\in\mathbb{R}^N$, and the symmetric group $S_N$ acts on $\mathbb{R}^N$ by permuting coordinates: $(\sigma x)_{\sigma(j)}:=x_j$. We call $x$ \emph{dominant} if its coordinates are weakly decreasing:
$x_1\geq x_2\geq\cdots \geq x_N$. Below we will often identify a partition $\lambda\in\Lambda^{(N)}$ with the dominant vector
$(\lambda_1,\lambda_2,\ldots,\lambda_N)\in\mathbb{Z}^N$.
For $x\in\mathbb{R}^N $ we write $\bar{x}:=x-\tfrac{|x|}{N}\mathbf{1}$ with $\mathbf{1}:=(1,\ldots ,1)\in\mathbb{Z}^N$ for its orthogonal projection onto $\mathbb{R}^N_0$
(cf.\ Remark \ref{hyperplane:rem}).
Intersecting $\mathbb{Z}^N\subset\mathbb{R}^N$ with this hyperplane or projecting it onto it, we obtain the root lattice $Q$ and the weight lattice $\bar{P}$ associated with $S_N$, respectively:
\begin{equation}\label{lattices}
Q:=\{\alpha \in\mathbb{Z}^N\mid |\alpha|=0\}\quad\text{and}\quad \bar{P}:=\{\bar{x}\mid x\in\mathbb{Z}^N\} .
\end{equation}
These lattices are dual to each other with respect to $\langle\cdot,\cdot\rangle$, and $Q\subset\bar{P}$ has index $[\bar{P}:Q]=N$.

\subsection{Preliminaries}
Two elementary consequences of \eqref{HL}, \eqref{C} are the translational symmetry of the Hall--Littlewood polynomials
\begin{subequations}
\begin{equation}\label{shift}
\forall \lambda\in \Lambda^{(N)}\ \text{and} \ c\in\mathbb{N}_0:\quad R_{\lambda+c^N}(z)=(z_1\cdots z_N)^c R_\lambda(z) ,
\end{equation}
and the identification of the constant Hall--Littlewood polynomial as the Poincar\'e polynomial of the symmetric group $S_N$ \cite[Chapter III, Eq.~(1.4)]{mac:symmetric}:
\begin{equation}\label{empty}
 R_{\emptyset}(z)=\sum_{\sigma\in S_N}C(z_{\sigma(1)},\ldots ,z_{\sigma(N)})=[N]!\, ,
\end{equation}
\end{subequations}
where $\emptyset$ denotes the empty partition (with all parts equal to $0$).

For any Laurent series $F(z)=\sum_{\beta\in\mathbb{Z}^N}F_\beta z^\beta$ that converges absolutely on the unit torus $\mathbb{T}:=\{z\in\mathbb{C}^N \mid |z_1|=\cdots =|z_N|=1\}$,
let us define the \emph{mean} of $F(z)$ as
\begin{subequations}
\begin{equation}\label{CT}
\Bigl\langle F(z) \Bigr\rangle :=\sum_{c\in\mathbb{Z}} F_{c\mathbf{1}} .
\end{equation}
Here $z^\beta:=z_1^{\beta_1}\cdots z_N^{\beta_N}$ denotes a monomial of total degree $|\beta|$, and we have in particular:
\begin{equation}\label{mean}
\Bigl\langle z^\beta\Bigr\rangle=\begin{cases}1&\text{if}\ \beta\in\mathbb{Z}\mathbf{1},\\ 0&\text{otherwise}.\end{cases}
\end{equation}
\end{subequations}
In other words, $\langle F(z)\rangle$ is the mean value of $F(e^{\mathrm{i}\xi})$ over the torus $\mathbb{R}^N_0/2\pi Q$, since the restriction of $z^\beta$ to $z=e^{\mathrm{i}\xi}$, $\xi\in\mathbb{R}^N_0$, is a character of this torus that is trivial iff $\beta\in\mathbb{Z}\mathbf{1}$.
Notice that for $|t|<1$ the rational functions $1/C(z^{\pm 1})$ (and hence $\Delta(z)$), $R_\lambda (z^{\pm 1})$, and the twisting factor $E_\alpha(z)$ \eqref{Ep} below are regular in $z$ within a neighborhood of $\mathbb{T}$, whence we can
identify any polynomial expression built from these building blocks with its absolutely convergent Laurent expansion.
The following lemma recasts the standard orthogonality relation for the Hall--Littlewood polynomials from \eqref{HL-orthogonality}, \eqref{wfs} in its frequently used
constant term formulation.

\begin{lemma}[Standard Orthogonality Relation for Hall--Littlewood Polynomials]\label{Mac-T:lem}
For $\lambda,\kappa\in\Lambda^{(n)}$:
\begin{equation}
\frac{1}{N!} \Bigl\langle R_\lambda(z)\,R_\kappa(z^{-1})\,\Delta(z) \Bigr\rangle = \frac{\delta_{\lambda,\kappa}}{\hat{\Delta}_\lambda}.
\end{equation}
\end{lemma}
\begin{proof}
Since $\Delta(z)$ is homogeneous of degree $0$ in $z$, every monomial $z^\beta$ in the expansion of $F(z):=R_\lambda(z)R_\kappa(z^{-1})\Delta(z)$ has degree $|\beta|=|\lambda|-|\kappa|$.
So if
$N$ does not divide $|\lambda|-|\kappa|$ (so in particular $\lambda\neq\kappa$), then $\langle F(z) \rangle=0$ by \eqref{mean}. If, on the other hand, $|\lambda|-|\kappa|=cN$ with (say) $c\geq 0$, then
\begin{equation*}
\langle F(z) \rangle=F_{c\mathbf{1}}= \frac{1}{(2\pi)^N} \int_{[0,2\pi]^N} e^{-c\mathrm{i}(\xi_1+\cdots+\xi_N)} F(e^{\mathrm{i}\xi})\, \mathrm{d}\xi=
\frac{ N!\, \delta_{\lambda,\kappa+c^N}}{\hat{\Delta}_\lambda}
\end{equation*}
(where the last equality relies on \eqref{shift} and \eqref{HL-orthogonality}). For $\lambda,\kappa\in\Lambda^{(n)}$ one has that
$\lambda_N=\kappa_N=0$, whence $\lambda=\kappa+c^N$ $\Longleftrightarrow$ $c=0$ and $\lambda=\kappa$.
The case $c<0$ follows similarly with the roles of $\lambda$ and $\kappa$ interchanged.
\end{proof}

\subsection{The Bethe map and the Poisson summation formula}
Let $\Phi:\mathbb{R}^N_0\to\mathbb{R}^N_0$ be given by
\begin{equation}\label{Phi}
\Phi(\xi)_j:=\frac{1}{2\pi}\Bigl(m\xi_j+\sum_{\substack{1\leq k\leq N\\ k\neq j}}v(\xi_j-\xi_k)\Bigr)\qquad (1\leq j\leq N),
\end{equation}
cf.\ Remark \ref{hyperplane:rem}, and let $\delta:=(N-1,N-2,\ldots ,1,0)$, so that $\bar\delta=\rho$. The critical equations \eqref{crit} read $\Phi(\xi)=\overline{\delta+\mu}$, i.e.
\begin{equation}\label{node}
\xi_\mu=\Phi^{-1}\bigl(\overline{\delta+\mu}\bigr)\qquad(\mu\in\Lambda^{(n,m)}),
\end{equation}
where $\Phi^{-1}$ makes sense by part (i) of the following lemma. Throughout we put $\boxed{L:=m+N}$, and we call $x\in\mathbb{Z}^N$ \emph{$L$-regular} if $x_j\not\equiv x_k\pmod L$ for all $j\neq k$.

The following lemma shows that the Bethe map $\Phi$ is a diffeomorphism of $\mathbb{R}_0^N$
that intertwines two actions of the affine permutation group $S_N\ltimes Q$: the action at
level $2\pi$, given by $\xi\mapsto\sigma\xi+2\pi\alpha$, on its domain, and the action at
level $L$, given by $x\mapsto\sigma x+L\alpha$, on its range ($\sigma\in S_N$, $\alpha\in Q$).
In this picture, the lemma also describes the preimages under $\Phi$ of the projected integral
points $\bar{x}$, $x\in\mathbb{Z}^N$: if $x$ fails to be $L$-regular, then
$\Phi^{-1}(\bar{x})$ belongs to the zero locus of the weight function $\Delta(e^{\mathrm{i}\xi})$
\eqref{wfs}, whereas if $x$ is $L$-regular, then $\Phi^{-1}(\bar{x})$ lies in the orbit of
precisely one Bethe root $\xi_\nu$, $\nu\in\Lambda^{(n,m)}$, under the action of
$S_N\ltimes Q$ at level $2\pi$.

\begin{lemma}[Bethe Map]\label{Phi:lem}
The map $\Phi$ \eqref{Phi} enjoys the following properties.
\begin{enumerate}
\item[(i)] $\Phi$ is a diffeomorphism of $\mathbb{R}^N_0$, and its Jacobian is given by $\det D\Phi(\xi)=\det H(\xi)/\bigl(m(2\pi)^n\bigr)$.
\item[(ii)] $\Phi(\sigma\xi)=\sigma\Phi(\xi)$ for $\sigma\in S_N$, and $\Phi(\xi+2\pi \alpha)=\Phi(\xi)+L\alpha$ for $\alpha\in Q$.
\item[(iii)] For $x\in\mathbb{Z}^N$, one has $\Delta(e^{\mathrm{i}\xi})=0$ at $\xi=\Phi^{-1}(\bar{x})$ iff $x$ is not $L$-regular.
\item[(iv)] For every $L$-regular $x\in\mathbb{Z}^N$ there is a unique pair $(\sigma,\nu)\in S_N\times\Lambda^{(n,m)}$ with $x\in\sigma(\delta+\nu)+\mathbb{Z}\mathbf{1}+LQ$, and thus $\Phi^{-1}(\bar x)\in\sigma\xi_\nu+2\pi Q$.
\end{enumerate}
\end{lemma}
\begin{proof}
(i) This property of $\Phi$ synthesizes the main point of the analysis of the Bethe Ansatz equations in \cite[Section 4]{die:diagonalization}. Specifically,
$2\pi\Phi$ is the gradient map, restricted to $\mathbb{R}^N_0$, of the Morse function $U(\xi):=\sum_{j<k}\int_0^{\xi_j-\xi_k}v(\theta)\mathrm{d}\theta+\frac{m}{2}\langle\xi,\xi\rangle$, whose Hessian is the matrix $H$ \eqref{Hess}. The Jacobi matrix $D\Phi=\frac{1}{2\pi}H|_{\mathbb{R}^N_0}$ is thus positive definite, because $H=m\,\mathrm{Id}+\sum_{j<k}v'(\xi_j-\xi_k)(e_j-e_k)(e_j-e_k)^{T}$ with $v'>0$ \eqref{vprime} (where the column vectors $e_1,\ldots ,e_N$ refer to the standard basis of $\mathbb{R}^N$). Since $v$ is odd and increasing, $\int_0^{\theta} v(\vartheta)\mathrm{d}\vartheta\geq 0$ for all
$\theta\in\mathbb{R}$, so for any $b\in\mathbb{R}^N_0$ the function $U(\xi)-2\pi\langle b,\xi\rangle$ (which is strictly convex
since its Hessian is $H$) satisfies $U(\xi)-2\pi\langle b,\xi\rangle\geq\frac{m}{2}\langle\xi,\xi\rangle-2\pi\langle b,b\rangle^{1/2}\langle\xi,\xi\rangle^{1/2}$
and is therefore also radially unbounded on $\mathbb{R}^N_0$;
hence it has exactly one critical point.
 In other words, $\Phi(\xi)=b$ has exactly one solution, and $\Phi$ is thus a diffeomorphism (by the inverse function theorem). Finally, $H$ is symmetric with $H\mathbf{1}=m\mathbf{1}$ by \eqref{Hess}, so $H$ preserves $\mathbb{R}^N_0=\mathbf{1}^\perp$ and $\det H=m\det(H|_{\mathbb{R}^N_0})=m(2\pi)^n\det D\Phi$.

(ii) The first statement is immediate from \eqref{Phi} by relabeling the summation index, since $(\sigma\xi)_{\sigma(j)}=\xi_j$. To check the quasi-periodicity, one observes that the integrand in \eqref{v} is $2\pi$-periodic with integral $2\pi$ over a period, so $v(\theta+2\pi k)=v(\theta)+2\pi k$ for $k\in\mathbb{Z}$. Hence, for $\alpha\in Q$: $\sum_{1\leq k\leq N,\, k\neq j}v(\xi_j-\xi_k+2\pi(\alpha_j-\alpha_k))=\sum_{1\leq k\leq N,\, k\neq j}v(\xi_j-\xi_k)+2\pi(N\alpha_j-|\alpha|)=\sum_{1\leq k\leq N,\, k\neq j}v(\xi_j-\xi_k)+2\pi N\alpha_j$.

(iii) Let $x_j-x_k=Lr$ with $j\neq k$, $r\in\mathbb{Z}$, let $\tau\in S_N$ be the transposition of $j$ and $k$, and set $\alpha:=r(e_j-e_k)\in Q$. Then $\tau x+L\alpha=x$, so $\overline{\tau x}+L\alpha=\bar{x}$ and, by (ii), $\Phi(\tau\xi+2\pi \alpha)=\tau\bar{x}+L\alpha=\Phi(\xi)$ for $\xi:=\Phi^{-1}(\bar{x})$. Hence $\tau\xi+2\pi \alpha=\xi$, i.e.\ $\xi_j-\xi_k=2\pi r$, so the factor $1-e^{\mathrm{i}(\xi_j-\xi_k)}$ of $\Delta(e^{\mathrm{i}\xi})$ vanishes. Conversely, if $\Delta(e^{\mathrm{i}\xi})=0$ for $\xi:=\Phi^{-1}(\bar{x})$, then $\xi_j-\xi_k=2\pi r$ for some $j\neq k$ and $r\in\mathbb{Z}$, and with $\tau$, $\alpha$ as before we get $\tau\xi+2\pi \alpha=\xi$. Applying $\Phi$ and using (ii) gives $\overline{\tau x}+L\alpha=\bar{x}$, i.e.\ $\tau x+L\alpha-x\in\mathbb{Z}\mathbf{1}$. The $j$th and $k$th components of this vector are $x_k-x_j+Lr$ and $x_j-x_k-Lr$, which, being equal, must both vanish. Hence $x_j-x_k=Lr$, so $x$ fails to be $L$-regular.

(iv) The vectors $\sigma(\delta+\nu)$, $(\sigma,\nu)\in S_N\times\Lambda^{(n,m)}$, are precisely the vectors $y\in\mathbb{Z}^N$ with distinct entries in $\{0,\ldots ,L-1\}$ one of which is $0$ (sort the entries of $y$ decreasingly as $L>y_{\sigma(1)}>\cdots>y_{\sigma(N)}=0$ and put $\nu_j:=y_{\sigma(j)}-(N-j)$). Moreover, each such $y$ determines $(\sigma,\nu)$. Given an $L$-regular $x$, it suffices to show that there is exactly one $c\in\mathbb{Z}$ for which $y:=x-c\mathbf{1}-L\alpha$ is of this form for some $\alpha\in Q$. The entries of $y$ lie in $\{0,\ldots ,L-1\}$ iff $\alpha_j=\lfloor (x_j-c)/L\rfloor$ for all $j$, and then $\alpha\in Q$ iff $f(c)=0$, where $f(c):=\sum_{j=1}^N\lfloor (x_j-c)/L\rfloor$ is a nonincreasing function of $c\in\mathbb{Z}$. The fact that the residues $x_j\bmod L$ are all distinct implies that $f(c)$ drops by exactly $1$ when $c$ passes from $x_j-Lk$ to $x_j-Lk+1$ and is constant otherwise. The quasi-periodicity $f(c-L)=f(c)+N$ then guarantees that $\{c\mid f(c)=0\}$ is a nonempty finite interval of consecutive integers. Since $y_j$ is the remainder of $x_j-c$ modulo $L$, it is clear that $y$ has a vanishing entry iff $c\equiv x_j\pmod L$ for some $j$, i.e.\ iff $f(c+1)=f(c)-1$, i.e.\ iff $c$ is the right endpoint of this interval. The last assertion follows in turn from \eqref{node} and (ii), since $\overline{\sigma(\delta+\nu)+c\mathbf{1}+L\alpha}=\sigma (\overline{\delta+\nu} )+L\alpha$.
\end{proof}

The following proposition expresses the left-hand side of the affine orthogonality relation in
Theorem \ref{affine-HL-orthogonality:thm} as a series, indexed by the root lattice $Q$, of
twisted versions of the mean in the standard orthogonality relation of
Lemma \ref{Mac-T:lem}; the term $\alpha=0$ reproduces the untwisted mean itself.
The identity in question results from the Poisson summation formula on the torus $\mathbb{R}^N_0/LQ$, pulled back to
$\mathbb{R}^N_0/2\pi Q$ by means of the Bethe map $\Phi$.

Here and below, $\mathrm{d}x$ and $\mathrm{d}\xi$ in integrals over (quotients of) the hyperplane
$\mathbb{R}^N_0\cong\mathbb{R}^n$ denote the Lebesgue measure on $\mathbb{R}^N_0$ induced by the
inner product $\langle\cdot,\cdot\rangle$, and $\operatorname{vol}(\mathbb{R}^N_0/\Gamma)$ denotes
the corresponding covolume of a lattice $\Gamma\subset\mathbb{R}^N_0$. Accordingly, the Jacobi
matrix $D\Phi$ of the Bethe map is regarded as an endomorphism of $\mathbb{R}^N_0$, and
$\det D\Phi$ denotes its determinant as such, so that the change of variables $x=\Phi(\xi)$ gives
$\mathrm{d}x=\det D\Phi\,\mathrm{d}\xi$ (cf.\ the relation $\det H=m(2\pi)^n\det D\Phi$ in the proof
of Lemma \ref{Phi:lem} (i)).

\begin{proposition}[Poisson Summation Expansion]\label{poisson:prp}
For $\lambda,\kappa\in\Lambda^{(n,m)}$:
\begin{subequations}
\begin{equation}\label{poisson}
\sum_{\nu\in\Lambda^{(n,m)}} R_\lambda( e^{\mathrm{i} \xi_\nu})   R_\kappa( e^{-\mathrm{i} \xi_\nu}) \Delta^{\mathrm{aff}} (\xi_\nu)
=\frac{1}{N!}\sum_{\alpha\in Q}\Bigl\langle R_\lambda(z)\,R_\kappa(z^{-1})\,\Delta(z)\,E_\alpha(z)\Bigr\rangle ,
\end{equation}
where
\begin{equation}\label{Ep}
E_\alpha(z):=z^{-L\alpha}A_\alpha(z)\quad \text{with} \quad A_\alpha(z):=\prod_{1\leq j<k\leq N}\Bigl(\frac{1-tz_jz_k^{-1}}{1-tz_kz_j^{-1}}\Bigr)^{\alpha_j-\alpha_k} .
\end{equation}
Here the sum on the right-hand side
is absolutely convergent, and for $\xi\in\mathbb{R}^N_0$ and $\sigma\in S_N$ one has that
\begin{equation}\label{Ep-Bethe}
E_\alpha(e^{\mathrm{i}\xi})=e^{-2\pi\mathrm{i}\langle \alpha,\Phi(\xi)\rangle}\quad \text{and}\quad
E_{\sigma \alpha}(e^{\mathrm{i}\xi})=E_\alpha(e^{\mathrm{i}\sigma^{-1}(\xi)}).
\end{equation}
\end{subequations}
\end{proposition}

\begin{proof}
Let $g:\mathbb{R}_0^N\to\mathbb{C}$ and $h:\mathbb{R}_0^N\to\mathbb{C}$ denote the smooth functions defined by
\begin{equation*}
g(\xi):=R_\lambda(e^{\mathrm{i}\xi})R_\kappa(e^{-\mathrm{i}\xi})\Delta(e^{\mathrm{i}\xi})\quad \text{and}\quad h:=(g/\det D\Phi)\circ\Phi^{-1}.
\end{equation*}
Then $g$ and $h$ are invariant with respect to the natural actions of $S_N\ltimes 2\pi Q$ and $S_N\ltimes LQ$ on the respective variables, cf.\ Lemma \ref{Phi:lem} (i), (ii).
Notice at this point that the invariance of $h$ is inherited from $g$ because $\Phi(\sigma\xi)=\sigma\Phi(\xi)$ and $\Phi(\xi+2\pi\alpha)=\Phi(\xi)+L\alpha$, which give in particular that $D\Phi(\sigma\xi)=P_\sigma D\Phi(\xi)P_\sigma^{-1}$ (where $P_\sigma$ denotes the permutation matrix associated with $\sigma$) and $D\Phi(\xi+2\pi\alpha)=D\Phi(\xi)$, and thus $\det D\Phi(\sigma\xi+2\pi\alpha)=\det D\Phi(\xi)$.
The upshot is that $g$ and $h$ descend to smooth $S_N$-invariant functions on $\mathbb{R}^N_0/(2\pi Q)$ and $\mathbb{R}^N_0/(LQ)$, respectively.
The characters of the latter torus, which are given by $x\mapsto e^{2\pi\mathrm{i}\langle\zeta,x\rangle}$ with $\zeta\in L^{-1}\bar{P}$, are trivial on the finite subgroup $\bar{P}/LQ$ iff $\zeta\in Q$.

Upon representing the smooth function $h$ by its absolutely convergent Fourier series
\begin{equation*}
h(x)=\sum_{\zeta\in L^{-1}\bar P}\hat h_\zeta\,e^{2\pi\mathrm{i}\langle\zeta,x\rangle}\quad\text{with}\quad
\hat h_\zeta:=\frac{1}{\operatorname{vol}(\mathbb{R}^N_0/LQ)}\int_{\mathbb{R}^N_0/LQ}h(x)\,e^{-2\pi\mathrm{i}\langle\zeta,x\rangle}\,\mathrm{d}x ,
\end{equation*}
we sum over $x\in\bar P/LQ$, using that
\begin{equation*}
\sum_{x\in\bar P/LQ}e^{2\pi\mathrm{i}\langle\zeta,x\rangle}=
\begin{cases}|\bar P/LQ| & \text{if } \zeta\in Q,\\ 0&\text{otherwise,}\end{cases}
\end{equation*}
(because---by Schur orthogonality---a character of a finite abelian group sums to its order if it is trivial and to zero otherwise),
so as to arrive at the following Poisson summation formula:
\begin{align*}
\sum_{x\in\bar P/LQ}h(x) &=|\bar P/LQ|\sum_{\alpha\in Q}\hat h_\alpha 
=\frac{|\bar{P}/LQ|}{\operatorname{vol}(\mathbb{R}^N_0/LQ)}\sum_{\alpha\in Q}\int_{\mathbb{R}^N_0/LQ}h(x)e^{-2\pi\mathrm{i}\langle \alpha,x\rangle}\mathrm{d}x\\
&\stackrel{(a)}{=}\frac{1}{\operatorname{vol}(\mathbb{R}^N_0/\bar{P})}\sum_{\alpha\in Q}\int_{\mathbb{R}^N_0/2\pi Q}g(\xi)E_\alpha(e^{\mathrm{i}\xi})\,\mathrm{d}\xi \\
&\stackrel{(b)}{=} N(2\pi)^n\sum_{\alpha\in Q} \Bigl\langle R_\lambda(z)\,R_\kappa(z^{-1})\,\Delta(z)\,E_\alpha(z)\Bigr\rangle .
\end{align*}
To infer the equality (a), one
substitutes $x=\Phi(\xi)$ (so that $\mathrm{d}x=\det D\Phi\,\mathrm{d}\xi$), uses the identity
$
e^{-2\pi\mathrm{i}\langle\alpha,\Phi(\xi)\rangle}=E_\alpha(e^{\mathrm{i}\xi})
$ from the first half of \eqref{Ep-Bethe}, and invokes the index relation $|\bar{P}/LQ|=\operatorname{vol}(\mathbb{R}^N_0/LQ)/\operatorname{vol}(\mathbb{R}^N_0/\bar{P})$.
The equality (b) hinges on $g(\xi)=R_\lambda(z)R_\kappa(z^{-1})\Delta(z)$ at $z=e^{\mathrm{i}\xi}$, on the volume relations $\operatorname{vol}(\mathbb{R}^N_0/2\pi Q)=(2\pi)^n\operatorname{vol}(\mathbb{R}^N_0/Q)=N(2\pi)^n\operatorname{vol}(\mathbb{R}^N_0/\bar{P})$, together with the observation that---for a Laurent series $F(z)$ converging absolutely on $\mathbb{T}$---the mean value of $F(e^{\mathrm{i}\xi})$ over $\mathbb{R}^N_0/2\pi Q$ is $\langle F(z)\rangle$ (cf.\ \eqref{mean}).

On the other hand, 
a direct computation of the sum in question yields
\begin{equation*}
\sum_{x\in\bar{P}/LQ}h(x)\stackrel{(c)}{=}\sum_{x\in\mathbb{Z}^N/(\mathbb{Z}\mathbf{1}+LQ)}\frac{g}{\det D\Phi}\bigl(\Phi^{-1}(\bar{x})\bigr)
\stackrel{(d)}{=} m (2\pi)^n\, N!\sum_{\nu\in\Lambda^{(n,m)}}\frac{g(\xi_\nu)}{\det H(\xi_\nu)} .
\end{equation*}
Here we use for (c) that $\bar{x}\in LQ$ iff $x\in\mathbb{Z}\mathbf{1}+LQ$, and for (d) the following three facts:
the classes of the $x$ that fail to be $L$-regular contribute zero, since $g$ vanishes at the corresponding points $\Phi^{-1}(\bar{x})$ (Lemma \ref{Phi:lem} (iii));
the classes of the $L$-regular $x$ are in bijection with the pairs $(\sigma,\nu)\in S_N\times\Lambda^{(n,m)}$, with $\Phi^{-1}(\bar{x})\in\sigma\xi_\nu+2\pi Q$ (Lemma \ref{Phi:lem} (iv)), so that the $S_N\ltimes 2\pi Q$ invariance of $g$ and of $\det H$ produces the factor $N!$;
and $\det D\Phi=\det H/\bigl(m(2\pi)^n\bigr)$ (Lemma \ref{Phi:lem} (i)).

The asserted identity in
\eqref{poisson} is now immediate upon comparing the two expressions for the sum of $h$ over the group $\bar{P}/LQ$ just obtained, and recalling \eqref{wfs}, \eqref{wf-affine}.
 Finally, to infer the identity $E_\alpha(e^{\mathrm{i}\xi})=e^{-2\pi\mathrm{i}\langle\alpha,\Phi(\xi)\rangle}$, one observes that $2\pi\langle \alpha,\Phi(\xi)\rangle=m\langle \alpha,\xi\rangle+\sum_{j<k}(\alpha_j-\alpha_k)v(\xi_j-\xi_k)$ (because $v$ is odd) and that
\begin{equation*}
e^{\mathrm{i}v(\theta)}=\frac{e^{\mathrm{i}\theta}-t}{1-te^{\mathrm{i}\theta}}=e^{\mathrm{i}\theta}\,\frac{1-te^{-\mathrm{i}\theta}}{1-te^{\mathrm{i}\theta}}
\end{equation*}
(both sides equal $1$ at $\theta=0$; differentiate and compare with \eqref{vprime}). Hence at $z=e^{\mathrm{i}\xi}$:
\begin{equation*}
e^{-2\pi\mathrm{i}\langle\alpha,\Phi(\xi)\rangle}=z^{-m\alpha}\prod_{j<k}(z_jz_k^{-1})^{-(\alpha_j-\alpha_k)}A_\alpha(z)=z^{-L\alpha}A_\alpha(z)=E_\alpha(z), 
\end{equation*}
because
$
\prod_{1\leq j<k\leq N}(z_jz_k^{-1})^{\alpha_j-\alpha_k}=z^{N\alpha-|\alpha|\mathbf{1}}=z^{N\alpha} $.
The covariance $E_{\sigma \alpha}(e^{\mathrm{i}\xi})=E_\alpha(e^{\mathrm{i}\sigma^{-1}(\xi)})$ follows in turn via Lemma \ref{Phi:lem} (ii).
\end{proof}

\begin{remark}\label{t=0:rem}
For $t=0$ the Hall--Littlewood polynomials reduce to Schur polynomials and
$\Delta(z)=\prod_{1\leq j\neq k\leq N}(1-z_j^{-1}z_k)$, while the Bethe map becomes linear:
$\Phi(\xi)=\frac{L}{2\pi}\xi$. The Bethe roots are then the equidistant points
$\xi_\mu=\frac{2\pi}{L}\overline{\delta+\mu}$ and $\det H(\xi)=mL^n$, so that
$\Delta^{\mathrm{aff}}(\xi_\mu)=\Delta(e^{\mathrm{i}\xi_\mu})/(NL^n)$. In this situation
Theorem \ref{affine-HL-orthogonality:thm} is classical: it amounts to the
discrete orthogonality of the Schur polynomials at the elements of finite order, i.e.\ to the
unitarity of the Kac--Peterson $S$-matrix of $\widehat{\mathfrak{sl}}_N$ at level $m$
(cf.\ \cite[\S 13.8]{kac:infinite-dimensional} and \cite[\S 4.2]{die:finite-dimensional}), and
it coincides with the discrete orthogonality of the antisymmetrized exponentials on the
simplex in \cite[Theorem 4.5]{li-xu:discrete}. Proposition \ref{poisson:prp}, in turn,
reduces for $t=0$ (when $E_\alpha(z)=z^{-L\alpha}$) to the antisymmetrization of the finite
Poisson summation formula for the lattice pair $LQ\subset\bar P$ underlying the discrete
Fourier analysis on the lattice of type $A_n$ in \cite[Theorem 2.4]{li-xu:discrete}, and in
this case only the term $\alpha=0$ on the right-hand side of \eqref{poisson} survives (cf.\
Lemma \ref{surviving:lem} below, in which all terms with $\alpha\neq 0$ carry a factor
$t^{r^2}$). For $t\neq 0$ the terms with $\alpha\neq 0$ no longer vanish, and their evaluation
in Sections \ref{graded:sec} and \ref{completion:sec} gives rise to the affine dual weights $\hat\Delta^{\mathrm{aff}}_\lambda$.
\end{remark}

\subsection{Graded Laurent series}\label{graded:sec}
By Lemma \ref{Mac-T:lem}, the term with $\alpha=0$ on the right-hand side of \eqref{poisson} equals $\delta_{\lambda,\kappa}/\hat{\Delta}_\lambda$. It remains to evaluate the contributions from the terms with $\alpha\neq 0$, which involve the \emph{twisted weight functions} $\Delta(z)E_\alpha(z)$.
All but finitely many of these terms will turn out to vanish. We will now develop machinery to determine which terms are zero and to unravel the fine structure of the surviving terms.

To this end, given $\alpha\in Q$, we grade the terms of the Laurent series $F(z)=\sum_\beta F_\beta z^\beta$ with respect to the degree $\deg_\alpha z^\beta:=\langle\beta,\alpha\rangle$
(which depends only on $\beta$ modulo $\mathbb{Z}\mathbf{1}$ since $|\alpha|=0$). Specifically, the component of degree $d\in\mathbb{Z}$ is given by
\begin{equation*}
F(z)_{\alpha;d}:=\sum_{\langle\beta,\alpha\rangle=d}F_\beta z^\beta ,
\end{equation*}
and if $\deg_\alpha F(z):=\sup\{\langle\beta,\alpha\rangle\mid F_\beta\neq 0\}$ is finite, then the terms of highest degree are singled out by
\begin{equation*}
F(z)^{\mathrm{top}}_\alpha:=F(z)_{\alpha;\deg_\alpha F(z)} .
\end{equation*}
Clearly $\bigl(F(z)G(z)\bigr)_{\alpha;d_F+d_G}=F(z)^{\mathrm{top}}_\alpha G(z)^{\mathrm{top}}_\alpha$ and $\bigl(F(z)G(z)\bigr)_{\alpha;d}=0$ for $d>d_F+d_G$, provided $d_F:=\deg_\alpha F(z)$ and $d_G:=\deg_\alpha G(z)$ are finite. 
With this notation we have (cf.\ \eqref{mean}, \eqref{Ep}):
\begin{equation}\label{top-mean}
\Bigl\langle R_\lambda(z)R_\kappa(z^{-1})\Delta(z)E_\alpha(z)\Bigr\rangle=\Bigl\langle F(z)_{\alpha;L\langle \alpha,\alpha\rangle}\,z^{-L\alpha}\Bigr\rangle ,
\end{equation}
with $F(z):=R_\lambda(z)R_\kappa(z^{-1})\Delta(z)A_\alpha(z)$,
because all monomials $z^{L\alpha+c\mathbf{1}}$, $c\in\mathbb{Z}$, are of degree $L\langle \alpha,\alpha\rangle$.

For $\alpha\in Q$ we put $|\alpha|_j:=\alpha_1+\cdots +\alpha_j$ ($1\leq j\leq N$), and for dominant $\alpha$ we put
\begin{equation*}
M(\alpha):=\max_{1\leq j\leq N}|\alpha|_j
\end{equation*}
 (so $M(\alpha)\geq 1$ if $\alpha\neq 0$). A special role will be played by the dominant vectors that, up to permutation, label the potentially surviving terms:
\begin{equation}\label{pa}
\alpha^{(r)}:=(\underbrace{1,\ldots ,1}_{r},\underbrace{0,\ldots ,0}_{N-2r},\underbrace{-1,\ldots ,-1}_{r})\qquad (1\leq r\leq N/2),
\end{equation}
for which we write $I_+:=\{1,\ldots ,r\}$, $I_0:=\{r+1,\ldots ,N-r\}$ and $I_-:=\{N+1-r,\ldots ,N\}$ so as to single out the indices at which the components of $\alpha^{(r)}$ equal $1$, $0$ and $-1$, respectively. (Notice that $\alpha^{(r)}=\omega_r+\omega_{N-r}$, where $\omega_k:=\overline{e_1+\cdots+e_k}$ ($1\leq k<N$) are the fundamental weights of $\mathfrak{sl}(N,\mathbb{C})$.) Given a subset $I\subset\{1,\ldots ,N\}$, let us write $z_I:=(z_j)_{j\in I}$, $(z_I)^c:=\prod_{j\in I}z_j^c$, and $R_\mu(z_I)$, $\Delta_I(z_I)$ for the Hall--Littlewood polynomial \eqref{HL}, \eqref{C} and the weight function \eqref{wfs} in the $|I|$ variables $z_j$, $j\in I$ (where $\mu$ is a partition of length at most $|I|$; the formulae in Section \ref{sec2} correspond to $I=\{1,\ldots,N\}$). The orthogonality relation \eqref{HL-orthogonality} holds for any number of variables and will be used for such selections of variables to evaluate the term of \eqref{poisson} labeled by $\alpha^{(r)}$ (cf.\ \eqref{orbits} below). The following two lemmas make this possible by unraveling the structure of the terms in question.

\begin{lemma}[Twisted Weight Functions]\label{weight-top:lem}
Let $\alpha\in Q$ be dominant and let $\alpha^{(r)}$ be as in \eqref{pa} ($1\leq r\leq N/2$). Then one has
\begin{subequations}
\begin{equation}
\deg_\alpha\bigl(\Delta(z)A_\alpha(z)\bigr)=N\langle \alpha,\alpha\rangle
\end{equation}
and
\begin{equation}\label{weight-top}
\bigl(\Delta(z)A_{\alpha^{(r)}}(z)\bigr)^{\mathrm{top}}_{\alpha^{(r)}}=t^{r^2}\,z^{N\alpha^{(r)}}\,\Delta_{I_+}(z_{I_+})\Delta_{I_0}(z_{I_0})\Delta_{I_-}(z_{I_-}) .
\end{equation}
\end{subequations}
\end{lemma}
\begin{proof}
When stripped of its monomial factor $z^{-L\alpha}$, the twisted weight function $\Delta(z)E_\alpha(z)$ factorizes as follows (cf.\ \eqref{wfs}, \eqref{C} and \eqref{Ep}):
\begin{equation*}
\Delta(z)A_\alpha(z)=\prod_{1\leq j<k\leq N}\phi_{jk}(z),
\end{equation*}
where
\begin{equation*}
\phi_{jk}(z):=(1-z_jz_k^{-1})(1-z_kz_j^{-1})(1-tz_jz_k^{-1})^{c_{jk}-1}(1-tz_kz_j^{-1})^{-c_{jk}-1}
\end{equation*}
and $c_{jk}:=\alpha_j-\alpha_k\geq 0$. If $c_{jk}=0$, all monomials of $\phi_{jk}(z)$ have degree $0$. If $c_{jk}\geq 1$, then $(1-z_jz_k^{-1})(1-tz_jz_k^{-1})^{c_{jk}-1}$ is a polynomial in $z_jz_k^{-1}$ of degree $c_{jk}$ with leading coefficient $(-1)^{c_{jk}}t^{c_{jk}-1}$, while the expansion of $(1-z_kz_j^{-1})(1-tz_kz_j^{-1})^{-c_{jk}-1}$ in powers of $z_kz_j^{-1}$ is $1$ plus terms of negative degree. Hence, for $c_{jk}\geq 1$,
\begin{equation*}
\phi_{jk}(z)^{\mathrm{top}}_\alpha=(-1)^{c_{jk}}t^{c_{jk}-1}(z_jz_k^{-1})^{c_{jk}},
\end{equation*}
while for $c_{jk}=0$ the top part is $\phi_{jk}(z)$ itself; in either case $\deg_\alpha\phi_{jk}(z)=c_{jk}^2$. Summing the degrees of all factors yields the first formula of the lemma:
\begin{equation*}
\deg_\alpha\bigl(\Delta(z)A_\alpha(z)\bigr)=\sum_{j<k}(\alpha_j-\alpha_k)^2=N\langle \alpha,\alpha\rangle-|\alpha|^2=N\langle \alpha,\alpha\rangle .
\end{equation*}
Moreover, for $\alpha=\alpha^{(r)}$ one has
\begin{equation*}
c_{jk}=\begin{cases}
1&\text{for the $2r(N-2r)$ pairs with $(j,k)\in (I_+\times I_0)\cup (I_0\times I_-)$}, \\
2&\text{for the $r^2$ pairs in $I_+\times I_-$},\\
0&\text{otherwise},
\end{cases}
\end{equation*}
whence
\begin{align*}
\bigl(\Delta(z)A_{\alpha^{(r)}}(z)\bigr)^{\mathrm{top}}_{\alpha^{(r)}} &=t^{r^2}\prod_{j<k}(z_jz_k^{-1})^{\alpha_j-\alpha_k}\prod_{c_{jk}=0}\phi_{jk}(z)\\
&=t^{r^2}z^{N\alpha^{(r)}}\Delta_{I_+}(z_{I_+})\Delta_{I_0}(z_{I_0})\Delta_{I_-}(z_{I_-}) .
\end{align*}
\end{proof}

\begin{lemma}[Hall--Littlewood Polynomials]\label{HL-top:lem}
Let $\lambda\in\Lambda^{(n,m)}$, $A:=\mathrm{m}_m(\lambda)$, $B:=\mathrm{m}_0(\lambda)$, and let $\tbin{A}{r}:=\frac{[A]!}{[r]!\,[A-r]!}$ ($0\leq r\leq A$) denote the $t$-binomial coefficient.
\begin{subequations}
\begin{enumerate}
\item[(i)] For $\alpha\in Q$ dominant:
\begin{equation}
\deg_\alpha R_\lambda(z)\leq mM(\alpha)\quad\text{and}\quad \deg_\alpha R_\lambda(z^{-1})\leq mM(\alpha).
\end{equation}
\item[(ii)] Let $1\leq r\leq N/2$. If $r>\min(A,B)$, then
\begin{equation}
\deg_{\alpha^{(r)}} R_\lambda(z)<mr\quad\text{and}\quad \deg_{\alpha^{(r)}} R_\lambda(z^{-1})<mr.
\end{equation}
If $r\leq\min(A,B)$ (note that $\min(A,B)\leq N/2$, as $A+B\leq N$), then  $\deg_{\alpha^{(r)}} R_\lambda(z)=\deg_{\alpha^{(r)}} R_\lambda(z^{-1})=mr$ and
\begin{equation}\label{HL-top}
\begin{split}
R_\lambda(z)^{\mathrm{top}}_{\alpha^{(r)}}& =c_r(\lambda)\,(z_{I_+})^{m}R_\gamma(z_{I_0}),\\
R_\lambda(z^{-1})^{\mathrm{top}}_{\alpha^{(r)}}&=c_r(\lambda)\,(z_{I_-})^{-m}R_\gamma(z_{I_0}^{-1}) ,
\end{split}
\end{equation}
where $\gamma:=(\lambda_{r+1},\ldots ,\lambda_{N-r})$ (a partition of length $\leq N-2r$) and
\begin{equation}\label{cr}
c_r(\lambda):=\tbin{A}{r}\tbin{B}{r}[r]!^2.
\end{equation}
\end{enumerate}
\end{subequations}
\end{lemma}
\begin{proof}
Let $\alpha'\in Q$, $s>1$, and substitute $z_j\mapsto z_js^{\alpha'_j}$ in \eqref{HL}, \eqref{C}: the term labeled by $\sigma$ becomes
\begin{equation}\label{scaling}
s^{\langle\sigma\lambda,\alpha'\rangle}z^{\sigma\lambda}\prod_{1\leq j<k\leq N}f_s\bigl(\sigma(j),\sigma(k)\bigr)\quad\text{with}\quad f_s(j,k):=\frac{1-tz_j^{-1}z_ks^{\alpha'_k-\alpha'_j}}{1-z_j^{-1}z_ks^{\alpha'_k-\alpha'_j}} .
\end{equation}
For $s\to\infty$ (and fixed generic $z$), one has that $f_s(j,k)$ tends to $1$, to $t$, or to $\frac{1-tz_j^{-1}z_k}{1-z_j^{-1}z_k}$, depending on whether $\alpha'_j>\alpha'_k$, $\alpha'_j<\alpha'_k$, or $\alpha'_j=\alpha'_k$, respectively. Since $R_\lambda(zs^{\alpha'})=\sum_{d\in\mathbb{Z}} s^d R_\lambda(z)_{\alpha';d}$ (a sum with finitely many nonvanishing terms, because $R_\lambda(z)$ is a polynomial), it follows that
\begin{equation*}
\deg_{\alpha'} R_\lambda(z)\leq D:=\max_{\sigma\in S_N}\langle\sigma\lambda,\alpha'\rangle ,
\end{equation*}
and that $R_\lambda(z)_{\alpha';D}$ is the sum of the limits of the terms \eqref{scaling} with $\langle\sigma\lambda,\alpha'\rangle=D$, i.e.
\begin{equation}\label{top-component}
R_\lambda(z)_{\alpha';D}=\sum_{\substack{\sigma\in S_N \\ \langle\sigma\lambda,\alpha'\rangle=D }} t^{\ell_{\alpha'}(\sigma)}\,z^{\sigma\lambda}\prod_{\substack{1\leq j<k\leq N\\ \alpha'_{\sigma(j)}=\alpha'_{\sigma (k)}}}\frac{1-tz_{\sigma(j)}^{-1}z_{\sigma(k)}}{1-z_{\sigma(j)}^{-1}z_{\sigma(k)}} ,
\end{equation}
where $\ell_{\alpha'}(\sigma):=\bigl|\{(j,k)\mid 1\leq j<k\leq N,\ \alpha'_{\sigma(j)}<\alpha'_{\sigma(k)}\}\bigr|$.

(i) Let $\alpha'$ be any permutation of $\alpha$ or of $-\alpha$. Since $\lambda_N=0$, Abel summation gives $\langle\lambda,\alpha'\rangle=\sum_{j=1}^{N-1}(\lambda_j-\lambda_{j+1})|\alpha'|_j\leq\lambda_1\max_j|\alpha'|_j\leq mM(\alpha)$ (because a sum of $j$ entries of $\alpha$ is at most $|\alpha|_j$, and a sum of $j$ entries of $-\alpha$ is at most $-(\alpha_{N+1-j}+\cdots +\alpha_N)=|\alpha|_{N-j}$). Moreover, $\langle\sigma\lambda,\pm\alpha\rangle=\langle\lambda,\pm\sigma^{-1}\alpha\rangle\leq mM(\alpha)$ for all $\sigma\in S_N$ (by the previous estimate with $\alpha'=\pm\sigma^{-1}\alpha$), so the preceding paragraph (with $\alpha'=\alpha$, respectively $\alpha'=-\alpha$) yields $\deg_\alpha R_\lambda(z)\leq mM(\alpha)$ and $\deg_\alpha R_\lambda(z^{-1})=\deg_{-\alpha}R_\lambda(z)\leq mM(\alpha)$.

(ii) The top-degree component $R_\lambda(z^{-1})^{\mathrm{top}}_{\alpha^{(r)}}$ is obtained from $R_\lambda(z)^{\mathrm{top}}_{-\alpha^{(r)}}$ by substituting $z\mapsto z^{-1}$, so it suffices to treat $\alpha'\in\{\alpha^{(r)},-\alpha^{(r)}\}$ simultaneously. Put $I'_\pm:=\{j\mid \alpha'_j=\pm 1\}$ and $I'_0:=I_0$, and for $\sigma\in S_N$ let $S_\epsilon:=\sigma^{-1}(I'_\epsilon)$ ($\epsilon\in\{+,0,-\}$). Then $\langle\sigma\lambda,\alpha'\rangle=\sum_{j\in S_+}\lambda_j-\sum_{j\in S_-}\lambda_j\leq mr$, with equality iff $S_+\subset J_m:=\{1,\ldots ,A\}$ and $S_-\subset J_0:=\{N-B+1,\ldots ,N\}$, which is possible iff $r\leq\min(A,B)$. This proves the first assertion. In the remaining case $D=mr$, and \eqref{top-component} becomes
\begin{equation*}
R_\lambda(z)_{\alpha';mr}=\sum_{\substack{\sigma\in S_N \\ S_+\subset J_m,\ S_-\subset J_0}} t^{\ell_{\alpha'}(\sigma)}\,z^{\sigma\lambda}\prod_{\substack{1\leq j<k\leq N\\ \alpha'_{\sigma(j)}=\alpha'_{\sigma (k)}}}\frac{1-tz_{\sigma(j)}^{-1}z_{\sigma(k)}}{1-z_{\sigma(j)}^{-1}z_{\sigma(k)}} .
\end{equation*}
For such $\sigma$, the pairs $j<k$ with $\alpha'_{\sigma(j)}<\alpha'_{\sigma(k)}$ (counted by $\ell_{\alpha'}(\sigma)$) are the pairs with $k\in S_+$, $j\notin S_+$ and the pairs with $j\in S_-$, $k\notin S_-$ (note that $j\in S_-$, $k\in S_+$ is impossible as $J_m$ precedes $J_0$). Similarly, the pairs $j<k$ with $\alpha'_{\sigma(j)}=\alpha'_{\sigma(k)}$ are those with $j,k$ both in $S_+$, both in $S_-$, or both in $S_0$. Parametrizing $\sigma$ by $S_+$, $S_-$ and the bijections $\sigma|_{S_+}:S_+\to I'_+$, $\sigma|_{S_-}:S_-\to I'_-$, $\sigma|_{S_0}:S_0\to I'_0$, and summing over the three bijections with the aid of \eqref{HL}, \eqref{C}, \eqref{shift}, \eqref{empty} (the labels being $(\lambda_j)_{j\in S_+}=(m^r)$, $(\lambda_j)_{j\in S_-}=(0^r)$ and $(\lambda_j)_{j\in S_0}=\gamma$, the latter because $\lambda$ is constant on $J_m$ and on $J_0$), we obtain
\begin{align*}
R_\lambda(z)_{\alpha';mr}&=\Bigl(\sum_{\substack{S_+\subset J_m\\ |S_+|=r}}t^{\#\{j<k\mid j\notin S_+,\, k\in S_+\}}\Bigr)\Bigl(\sum_{\substack{S_-\subset J_0\\ |S_-|=r}}t^{\#\{j<k\mid j\in S_-,\, k\notin S_-\}}\Bigr)\\
&\qquad\times [r]!^2(z_{I'_+})^mR_\gamma(z_{I'_0}) .
\end{align*}
The two sums are the generating functions of the $r$-element subsets of the $A$-element set $J_m$, respectively of the $B$-element set $J_0$, by the number of inversions of the $0$-$1$ words given by their characteristic functions. Hence, the respective sums are given by the $t$-binomial coefficients $\tbin{A}{r}$ and $\tbin{B}{r}$, see, e.g.,~\cite[Proposition 1.7.1]{sta:enumerative}. The resulting polynomial is nonzero (as $R_\gamma\neq 0$ by \eqref{HL-orthogonality}), so $\deg_{\alpha'}R_\lambda(z)=mr$ and $R_\lambda(z)^{\mathrm{top}}_{\alpha'}=R_\lambda(z)_{\alpha';mr}$. For $\alpha'=\alpha^{(r)}$ one has $I'_\pm=I_\pm$, which gives the first formula in \eqref{HL-top}; for $\alpha'=-\alpha^{(r)}$ one has $I'_\pm=I_\mp$, and the substitution $z\mapsto z^{-1}$ gives the second.
\end{proof}

\subsection{Collecting surviving terms}\label{completion:sec}
Let $\lambda,\kappa\in\Lambda^{(n,m)}$; we write $A:=\mathrm{m}_m(\lambda)$, $B:=\mathrm{m}_0(\lambda)$, $A':=\mathrm{m}_m(\kappa)$, $B':=\mathrm{m}_0(\kappa)$, and adopt the notational conventions established above.
By the covariance $E_{\sigma \alpha}(e^{\mathrm{i}\xi})=E_\alpha(e^{\mathrm{i}\sigma^{-1}(\xi)})$ and the $S_N$-invariance of $R_\lambda(z)R_\kappa(z^{-1})\Delta(z)$ and of $\langle\cdot\rangle$ \eqref{CT}, \eqref{mean}, the terms on the right-hand side of the Poisson summation expansion
\eqref{poisson} are constant along $S_N$-orbits, so grouping them according to the dominant representatives gives
\begin{multline}\label{orbits}
\sum_{\nu\in\Lambda^{(n,m)}} R_\lambda( e^{\mathrm{i} \xi_\nu})   R_\kappa( e^{-\mathrm{i} \xi_\nu}) \Delta^{\mathrm{aff}} (\xi_\nu)
=\frac{\delta_{\lambda,\kappa}}{\hat{\Delta}_\lambda}\\
+\sum_{\substack{\alpha\in Q\setminus\{0\}\\ \text{dominant}}}\frac{1}{|\{\sigma\in S_N\mid \sigma \alpha=\alpha\}|}\Bigl\langle R_\lambda(z)\,R_\kappa(z^{-1})\,\Delta(z)\,E_\alpha(z)\Bigr\rangle ,
\end{multline}
where the term $\alpha=0$ was evaluated by Lemma \ref{Mac-T:lem}. The following lemma reveals, in particular, that only finitely many terms of \eqref{poisson} are nonzero.

\begin{lemma}[Surviving and Vanishing Terms]\label{surviving:lem}
For $\alpha\neq 0$ dominant:
\begin{align*}
&  \Bigl\langle R_\lambda(z)\,R_\kappa(z^{-1})\,\Delta(z)\,E_{\alpha}(z)\Bigr\rangle  =\\
& \left\{ 
            \begin{array}{ll}
   c_r(\lambda)c_r(\kappa)\,t^{r^2} \times \\
\Bigl\langle R_\gamma(z_{I_0})R_{\gamma'}(z_{I_0}^{-1})\Delta_{I_+}(z_{I_+})\Delta_{I_0}(z_{I_0})\Delta_{I_-}(z_{I_-})\Bigr\rangle
            &\substack{\text{if }\alpha=\alpha^{(r)}\text{ with}\\ r\leq\min(A,B,A',B')} , \\ [2.5ex] 
              0   & \text{otherwise},
            \end{array} 
          \right. 
 \end{align*} 
where $\gamma:=(\lambda_{r+1},\ldots ,\lambda_{N-r})$, $\gamma':=(\kappa_{r+1},\ldots ,\kappa_{N-r})$, and $c_r$ is given by \eqref{cr}.
\end{lemma}

\begin{proof}
For $F(z)$ as in \eqref{top-mean}, one has (cf.\ Lemmas \ref{weight-top:lem} and \ref{HL-top:lem} (i)):
\begin{equation*}
\deg_\alpha F(z)\leq 2mM(\alpha)+N\langle \alpha,\alpha\rangle.
\end{equation*}
 Let $p$ be the number of positive entries of $\alpha$, so
  \begin{equation*}
 M(\alpha)=|\alpha|_p=\sum_{1\leq j\leq p}\alpha_j=\sum_{p<j\leq N}|\alpha_j|. 
 \end{equation*}
 Since $\alpha_j^2\geq |\alpha_j|$ for integers, with equality iff $\alpha_j\in\{-1,0,1\}$, we get $\langle \alpha,\alpha\rangle\geq 2M(\alpha)$, with equality iff all entries of $\alpha$ lie in $\{-1,0,1\}$, i.e.\ (as $\alpha$ is dominant with $|\alpha|=0$) iff $\alpha=\alpha^{(r)}$ with $r=M(\alpha)$. If $\alpha$ is not of the form $\alpha^{(r)}$, then $\deg_\alpha F(z)\leq 2mM(\alpha)+N\langle \alpha,\alpha\rangle<L\langle \alpha,\alpha\rangle$, whence the pertinent mean vanishes by \eqref{top-mean}. For $\alpha=\alpha^{(r)}$ we have that $L\langle \alpha^{(r)},\alpha^{(r)}\rangle=2rL=2mr+2rN$, so by Lemma \ref{HL-top:lem} (ii) the mean vanishes unless $r\leq\min(A,B,A',B')$. In the latter case \eqref{top-mean}, \eqref{weight-top} and \eqref{HL-top} reveal
that $\bigl\langle R_\lambda(z)\,R_\kappa(z^{-1})\,\Delta(z)\,E_{\alpha^{(r)}}(z)\bigr\rangle$ equals
\begin{align*}
&\Bigl\langle R_\lambda(z)^{\mathrm{top}}_{\alpha^{(r)}}R_\kappa(z^{-1})^{\mathrm{top}}_{\alpha^{(r)}}\bigl(\Delta(z)A_{\alpha^{(r)}}(z)\bigr)^{\mathrm{top}}_{\alpha^{(r)}}z^{-L\alpha^{(r)}}\Bigr\rangle\\
&=c_r(\lambda)c_r(\kappa)\,t^{r^2}\Bigl\langle R_\gamma(z_{I_0})R_{\gamma'}(z_{I_0}^{-1})\Delta_{I_+}(z_{I_+})\Delta_{I_0}(z_{I_0})\Delta_{I_-}(z_{I_-})\Bigr\rangle ,
\end{align*}
where we used that the monomial $(z_{I_+})^m(z_{I_-})^{-m}z^{N\alpha^{(r)}}$ is equal to $z^{L\alpha^{(r)}}$. 
\end{proof}

The mean on the right-hand side of Lemma \ref{surviving:lem} has monomial exponents with vanishing sum over $I_+$ (and over $I_-$), so by \eqref{CT} only the constant term (the coefficient of $z^0$) contributes to this mean.
It therefore equals the normalized integral over the full torus $\mathbb{T}$, which factorizes over
the three blocks of variables (with $e^{\mathrm{i}\xi_I}:=(e^{\mathrm{i}\xi_j})_{j\in I}$). 
Specifically, by applying
\eqref{HL-orthogonality} in $r$ variables (with $\lambda=\kappa=\emptyset$ and $R_\emptyset=[r]!$) and in $N-2r$ variables (with $\lambda=\gamma$ and $\kappa=\gamma'$), the block integrals become:
\begin{gather*}
\frac{1}{(2\pi)^{r}}\int_{[0,2\pi]^{r}}\Delta_{I_\pm}(e^{\mathrm{i}\xi_{I_\pm}})\,\mathrm{d}\xi_{I_\pm}=\frac{r!}{[r]!}, \\
\frac{1}{(2\pi)^{N-2r}}\int_{[0,2\pi]^{N-2r}}R_\gamma(e^{\mathrm{i}\xi_{I_0}})R_{\gamma'}(e^{-\mathrm{i}\xi_{I_0}})\Delta_{I_0}(e^{\mathrm{i}\xi_{I_0}})\,\mathrm{d}\xi_{I_0}=\frac{(N-2r)!\,\delta_{\gamma,\gamma'}}{\hat{\Delta}_\gamma} ,
\end{gather*}
where the multiplicities in $\hat\Delta_\gamma$ are counted among the $N-2r$ parts $\gamma_1,\ldots ,\gamma_{N-2r}$. Hence $\mathrm{m}_m(\gamma)=A-r$, $\mathrm{m}_0(\gamma)=B-r$ and $\mathrm{m}_i(\gamma)=\mathrm{m}_i(\lambda)$ for $0<i<m$ (with the convention that $\gamma=\gamma'=\emptyset$ and $\hat\Delta_\gamma=1$ if $N=2r$, i.e.\ if the block $I_0$ is empty). Since $|\{\sigma\mid\sigma \alpha^{(r)}=\alpha^{(r)}\}|=r!^2(N-2r)!$, the term of \eqref{orbits} labeled by $\alpha^{(r)}$ is thus evaluated as:
\begin{equation*}
\frac{c_r(\lambda)c_r(\kappa)\,t^{r^2}}{[r]!^2}\,\frac{\delta_{\gamma,\gamma'}}{\hat{\Delta}_\gamma}
=\delta_{\lambda,\kappa}\,t^{r^2}\tbin{A}{r}^2\tbin{B}{r}^2\frac{[r]!^2[A-r]![B-r]!}{\hat{\Delta}_\lambda[A]![B]!}
=\frac{\delta_{\lambda,\kappa}}{\hat{\Delta}_\lambda}\,t^{r^2}\tbin{A}{r}\tbin{B}{r} ,
\end{equation*}
where we used that $\gamma=\gamma'$ iff $\lambda=\kappa$ (both being equal to $(m^r,\gamma,0^r)$ when $r\leq\min(A,B,A',B')$), as well as $\hat{\Delta}_\lambda/\hat{\Delta}_\gamma=[A-r]!\,[B-r]!/([A]!\,[B]!)$ \eqref{wfs}. Upon substituting in \eqref{orbits} and invoking the $q$-Chu--Vandermonde summation formula \cite[p.~190]{sta:enumerative}, we finally arrive at
\begin{align*}
\sum_{\nu\in\Lambda^{(n,m)}} R_\lambda( e^{\mathrm{i} \xi_\nu})   R_\kappa( e^{-\mathrm{i} \xi_\nu}) \Delta^{\mathrm{aff}} (\xi_\nu)
&=\frac{\delta_{\lambda,\kappa}}{\hat{\Delta}_\lambda}\sum_{r=0}^{\min(A,B)}t^{r^2}\tbin{A}{r}\tbin{B}{r}\\
&=\frac{\delta_{\lambda,\kappa}}{\hat{\Delta}_\lambda}\,\frac{[A+B]!}{[A]!\,[B]!}\stackrel{ \eqref{dual-wf-affine}}{=}\frac{\delta_{\lambda,\kappa}}{\hat{\Delta}^{\mathrm{aff}}_\lambda} ,
\end{align*}
which completes the proof of Theorem \ref{affine-HL-orthogonality:thm}.\qed

\bibliographystyle{amsplain}

\end{document}